\documentclass{mem}
\usepackage{natbib}\usepackage{txfonts}\usepackage{balance}
\usepackage{graphicx}
\usepackage[a4paper]{hyperref}
\idline{75}{282}
\begin{document}
\def\teff{$T\rm_{eff }$}
\def\logg{\mbox{log~{\it g}}}
\def\Msun{\mbox{$M$$_{\odot}$}}
\def\Msunb{\mbox{$M$$_{\odot}$} }
\def\rpro{\mbox{$r$-process}}
\def\spro{\mbox{$s$-process}}
\def\ncap{\mbox{$n$-capture}}

\title{
The Role of Primary $^{16}$O as a Neutron Poison in 
AGB Stars and Fluorine primary production at Halo Metallicities}

   \subtitle{}

\author{
R. \,Gallino\inst{1,3}, 
S. \,Bisterzo\inst{1}, 
S. \, Cristallo\inst{2,3},
\and O. \,Straniero\inst{3}
          }

  \offprints{R. Gallino}

\institute{
Dipartimento di Fisica Generale, 
Universit\`a di Torino, 10125 (To) Italy
\and
Departamento de Fisica Teorica y del Cosmos, 
Universidad de Granada, Campus de Fuentenueva, 
18071 Granada, Spain
\and
INAF-Osservatorio Astronomico di Collurania, 
via M. Maggini, 64100 Teramo, Italy\\
\email{gallino@ph.unito.it}
}

\authorrunning{Gallino et al.}

\titlerunning{Low metallicity AGBs: primary $^{16}$O and  $^{19}$F}

\abstract{
The discovery of a historical bug in the s-post-process AGB code 
obtained so far by the Torino group forced us to 
reconsider the role of primary $^{16}$O in the $^{13}$C-pocket, produced by the
$^{13}$C($\alpha$, n)$^{16}$O reaction, as important neutron poison for 
the build up of the s-elements at Halo metallicities.
The effect is noticeable only for the highest $^{13}$C-pocket efficiencies
(cases ST*2 and ST). For Galactic disc metallicities, the bug effect is negligible.
A comparative analysis of the neutron poison effect of other primary isotopes 
($^{12}$C, $^{22}$Ne and its progenies) is presented. 
The effect of proton captures, by  $^{14}$N(n, p)$^{14}$C, boosts a primary
production of Fluorine in Halo AGB stars, with [F/Fe] comparable to [C/Fe],
without affecting the s-elements production.  

}
\maketitle{}

\section{Discovery of a bug in the s-post-process code.}

We discovered an old bug in the s-post-process code that follows the
s-process production in AGB stars (Gallino et al. 1998, Straniero et al.
2003). In the $^{13}$C-pocket, where the major
neutron source $^{13}$C($\alpha$,n)$^{16}$O is activated, the poison effect
of $^{16}$O was overlooked.
Despite its very low neutron capture cross section, 
during the first $^{13}$C-pocket, the large primary $^{16}$O abundance makes the product 
$\sigma$$N$ to dominate over the other major light neutron poisons in reducing
the neutrons available for the Fe seeds. This is shown in
Table~\ref{poisons} for case ST (Gallino et al. 1988), AGB initial mass $M$ = 1.5 \Msun, 
[Fe/H] = $-$2.3. Note that the neutrons captured by the very abundant primary
$^{12}$C are almost fully recycled by $^{13}$C($\alpha$,n)$^{16}$O. 
In the computations, the efficiency of the $^{13}$C-pocket in the interpulse is kept
unchanged at every thermal pulse (TP) (see K{\"a}ppeler et al. (2010) and references
therein). In advanced TPs, the progressive addition in the He
intershell of primary $^{14}$N 
\footnote{by conversion if the H-burning shell of primary $^{12}$C mixed with
the envelope by previous third dredge up (TDU) episodes, followed by conversion to 
primary $^{22}$Ne by double $\alpha$ capture in the early
phase of the next TP.} 
makes $^{22}$Ne  and its progenies to become primary
isotopes and to largely overcome the poisoning effect of $^{16}$O.

\begin{table*}
\caption{Light neutron poisons in the first and in the last $^{13}$C-pocket
(case ST) for an AGB model of initial mass 1.5 \Msun, metallicity [Fe/H] = $-$2.3.}
\label{poisons}
\begin{center}
\begin{tabular}{lcccccc}
\hline
         &      $<$$\sigma$$>$ (mbarn) & 1$^{st}$ pocket& 1$^{st}$ pocket&last
pocket&last pocket\\
&  30 keV&Number fraction &$\sigma$N&Number fraction &$\sigma$N\\
\hline
$^{12}$C        &       0.0154 &    1.7$\times$10$^{-2}$ &
2.6$\times$10$^{-4}$& 1.6$\times$10$^{-2}$&
2.5$\times$10$^{-4}$\\
$^{16}$O        &       0.0380 &     1.8$\times$10$^{-3}$ &
6.8$\times$10$^{-5}$&
9.6$\times$10$^{-4}$&3.6$\times$10$^{-5}$\\
$^{22}$Ne&              0.059  &     3.6$\times$10$^{-6}$ &
2.1$\times$10$^{-7}$
&7.7$\times$10$^{-4}$&4.5$\times$10$^{-5}$\\
$^{25}$Mg       &       6.4    &     8.4$\times$10$^{-8}$ &
5.4$\times$10$^{-7}$&5.5$\times$10$^{-6}$ 
&3.5$\times$10$^{-5}$\\
\hline
$^{56}$Fe  &      11.7   &   1.1$\times$10$^{-7}$  & 1.3$\times$10$^{-6}$             
&6.6$\times$10$^{-8}$&7.7$\times$10$^{-7}$\\
\hline
\end{tabular}
\end{center}
\end{table*}

For high $^{13}$C-pocket efficiencies, cases ST*2 to ST, 
and metallicities [Fe/H] $<$ 2, there is a sensible decrease of the
s-element production in the envelope at the end of the AGB phase. This is shown
in Figure 1 for [hs/Fe] by comparing in the bottom panel the updated results with respect to previous 
results (top panel) in which the $^{16}$O produced by $\alpha$ capture on  $^{13}$C was not accounted for. 
For lower $^{13}$C-pocket 
efficiencies or for Galactic disk metallicities the differences are
negligible. The [ls/Fe] ratio changes in the same way, so that the
s-process indicator [hs/ls] is only marginally affected, as is evident in
Figure 2 by comparing the results of the top and bottom panels. Some differences are shown
in the second s-process indicator, [Pb/hs], as reported in Figure 3.
As indicated in Figure 4, a large spread of $^{13}$C-pocket efficiencies at any given metallicity
appears necessary to account for spectroscopic observations of s-process
enhanced stars. Spectroscopic observations of carbon-rich and s-process enhanced metal-poor stars
(CEMP-s) encompass  the range of theoretical predictions between ST/2 and
ST/50, averaging  around case ST/4. For Galactic
disk s-enriched stars the spectroscopic observations of different classes of
s-rich stars encompass the range of theoretical predictions of [hs/ls] between ST*2
and ST/3, averaging around the ST case. Spectroscopic observations are
discussed in the review by K{\"a}ppeler et al. (2010).
[Pb/hs] typically increases with decreasing metallicity.

\section{Primary production of Fluorine in low metallicity CEMP-s stars }   
% {figure*} con * 1 colonna, senza * 2 colonne
\begin{figure}[t!]
\resizebox{\hsize}{!}{\includegraphics[angle=-90]{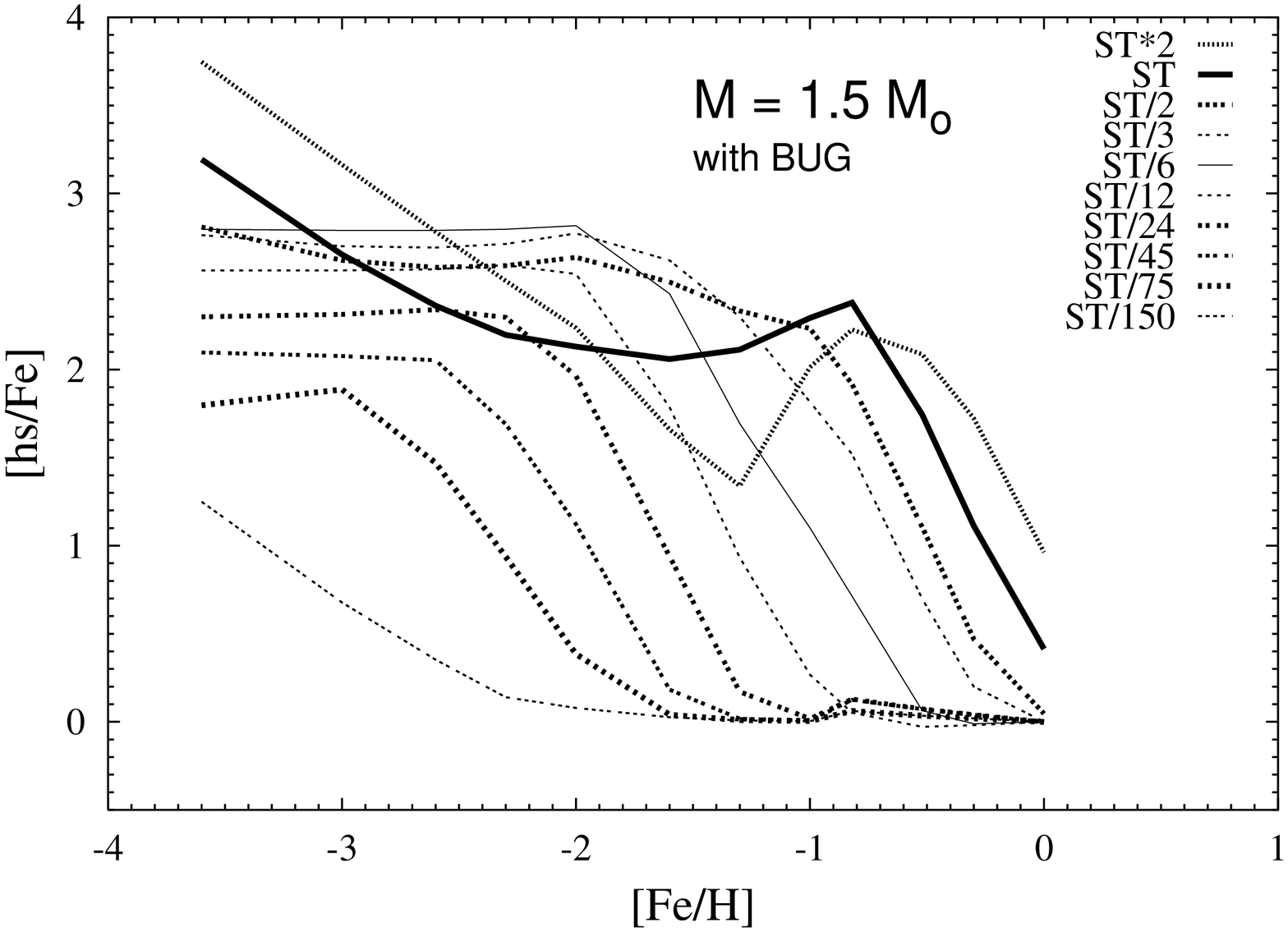}}
\resizebox{\hsize}{!}{\includegraphics[angle=-90]{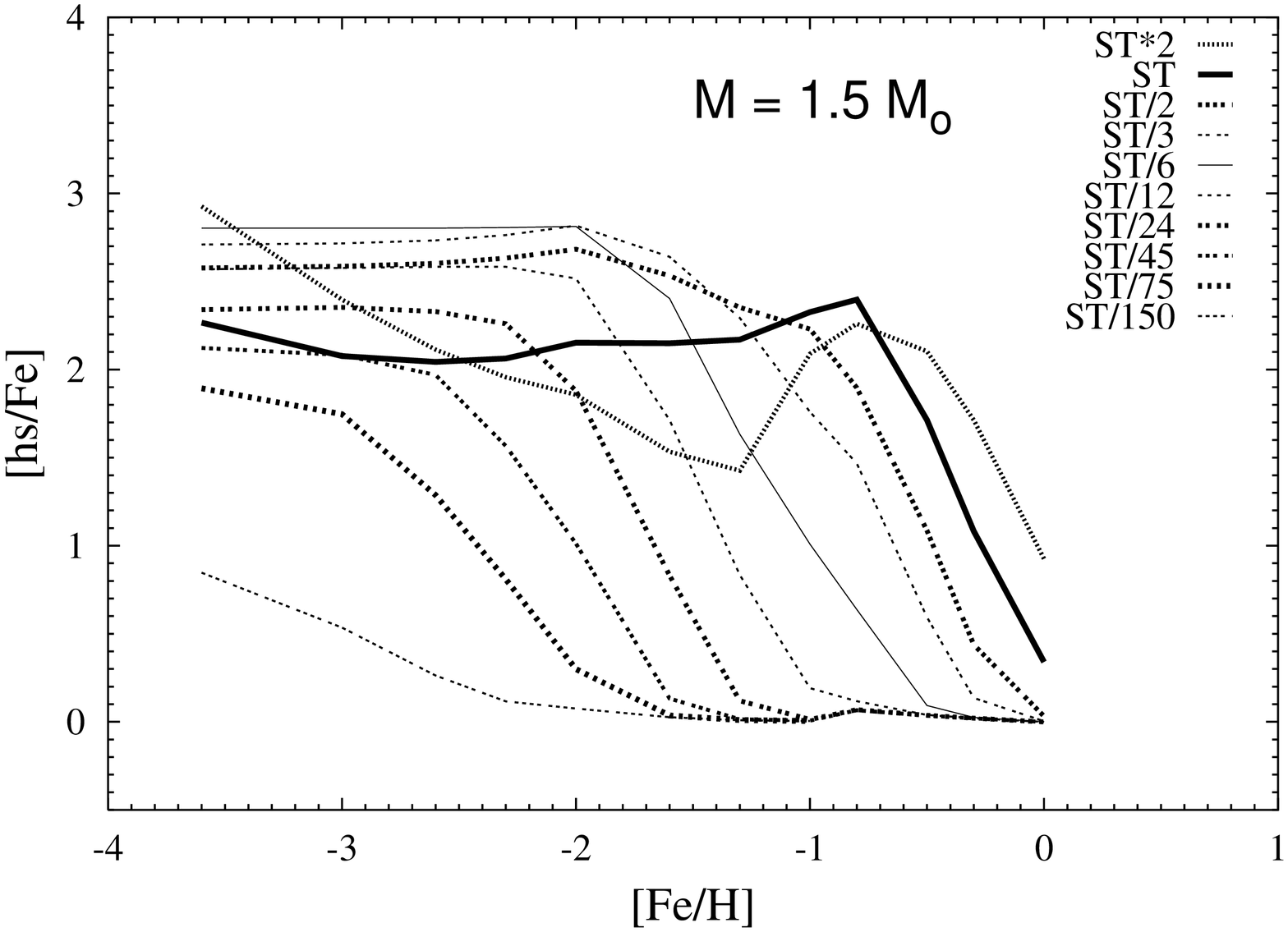}}
\caption{\footnotesize
The second s-peak [hs/Fe] for different $^{13}$C-pocket efficiencies. Top
panel: previous results with the $^{16}$O bug in the pocket; bottom panel;
updated results. 
}
\label{bugnobug}
\end{figure}

% {figure*} con * 1 colonna, senza * 2 colonne
\begin{figure}[t!]
\resizebox{\hsize}{!}{\includegraphics[angle=-90]{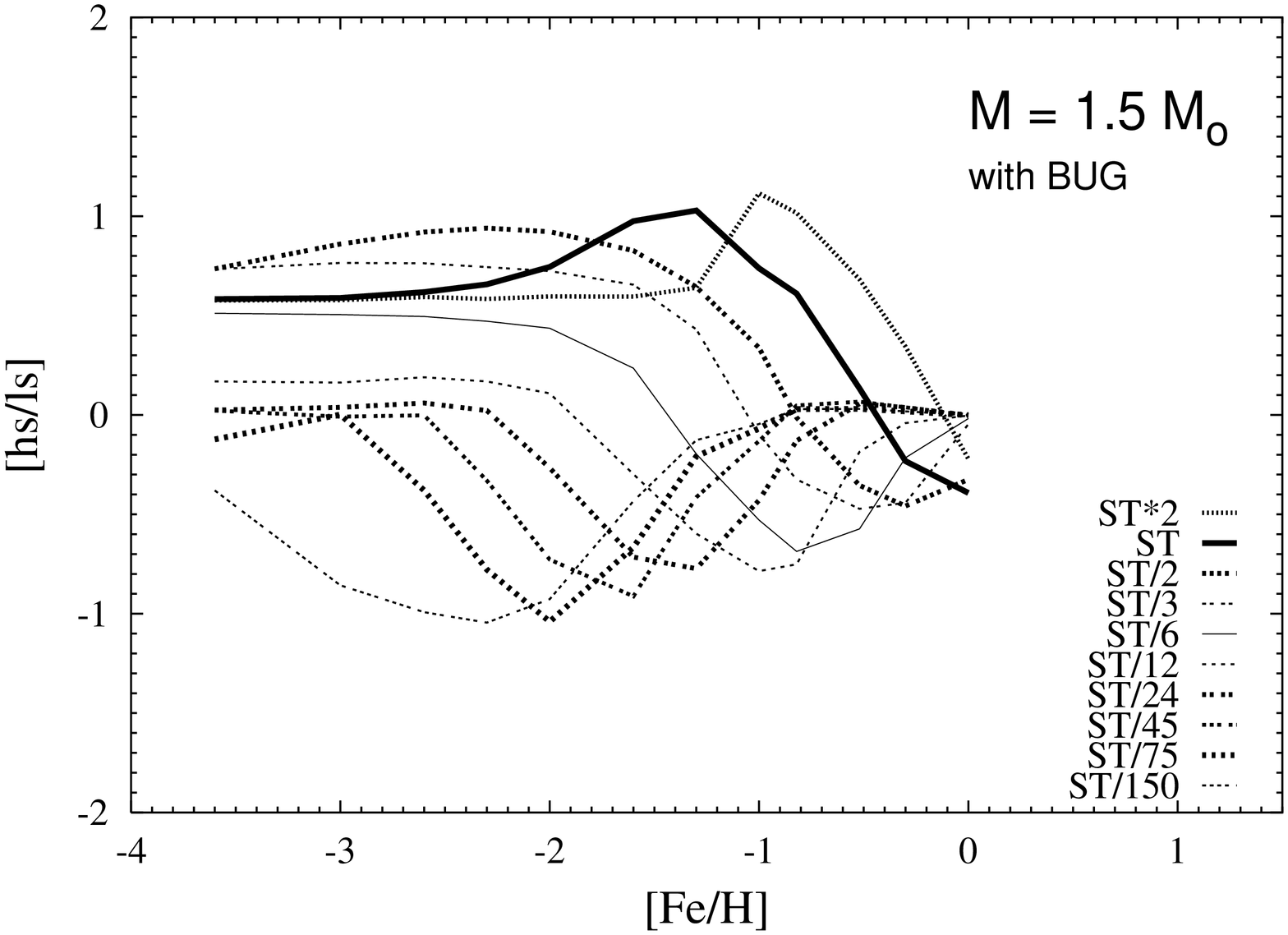}}
\resizebox{\hsize}{!}{\includegraphics[angle=-90]{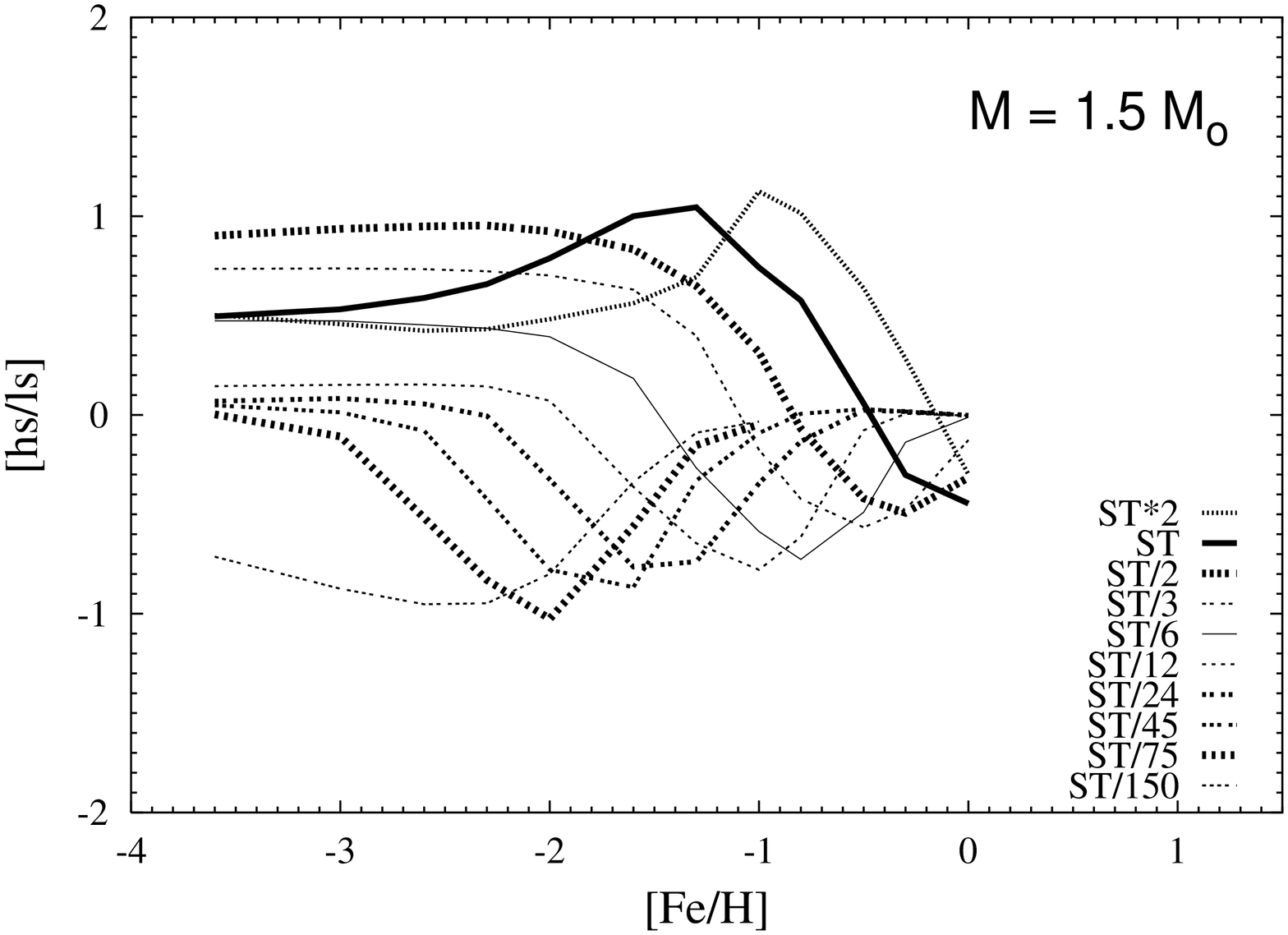}}
\caption{\footnotesize
The first s-process indicator [hs/ls] for different $^{13}$C-pocket efficiencies. Top   
panel: previous results with the $^{16}$O bug in the pocket; bottom panel:  
updated results.
}
\label{bugnobug1}
\end{figure}

% {figure*} con * 1 colonna, senza * 2 colonne
\begin{figure}[t!]
\resizebox{\hsize}{!}{\includegraphics[angle=-90]{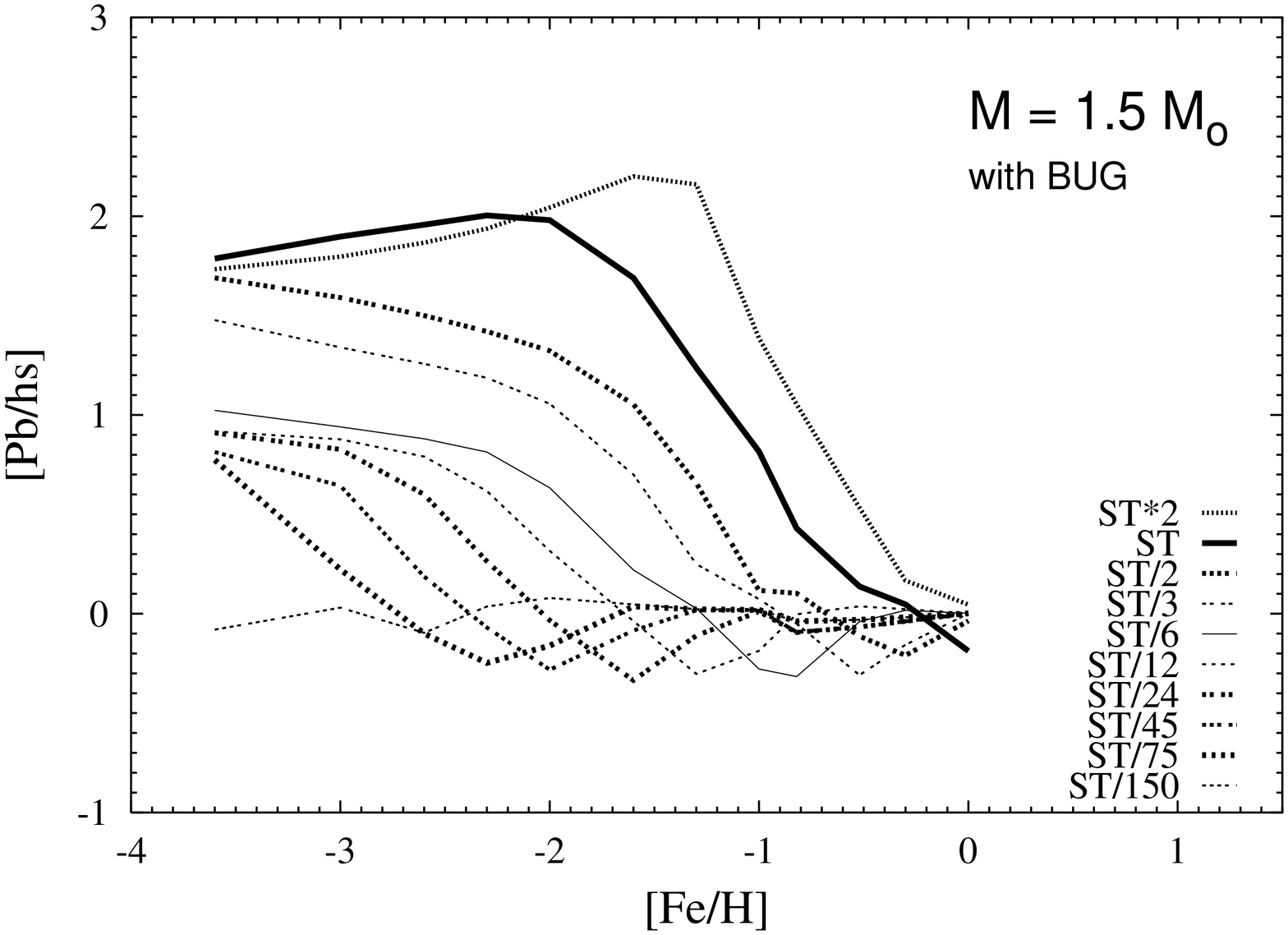}}
\resizebox{\hsize}{!}{\includegraphics[angle=-90]{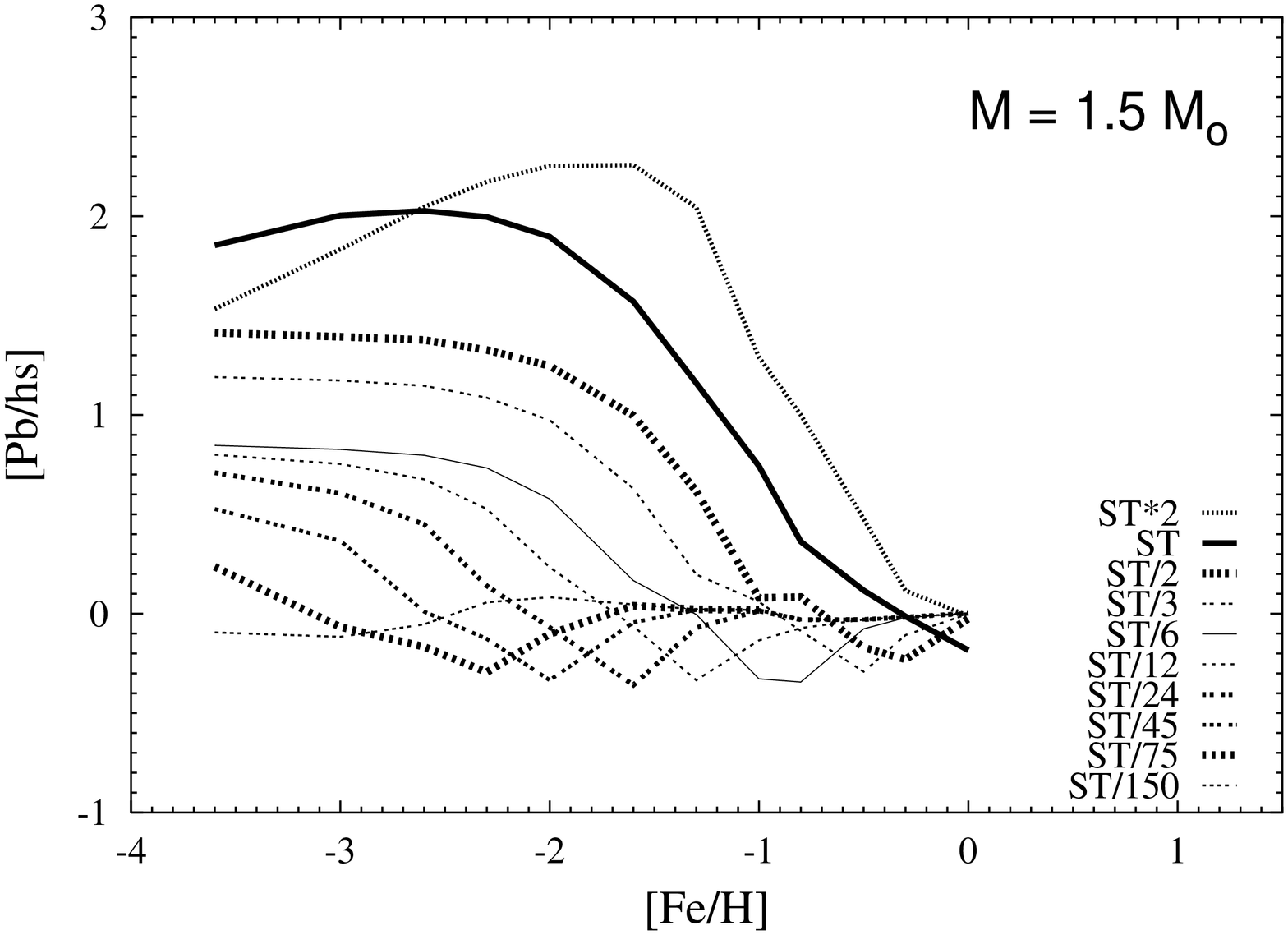}}
\caption{\footnotesize
The second s-process indicator [Pb/hs] for different $^{13}$C-pocket
efficiencies. Top
panel: previous results with the $^{16}$O bug in the pocket; bottom panel;  
updated results.
}
\label{bugnobug2}
\end{figure}

\begin{figure}[t!]
\resizebox{\hsize}{!}{\includegraphics[angle=-90]{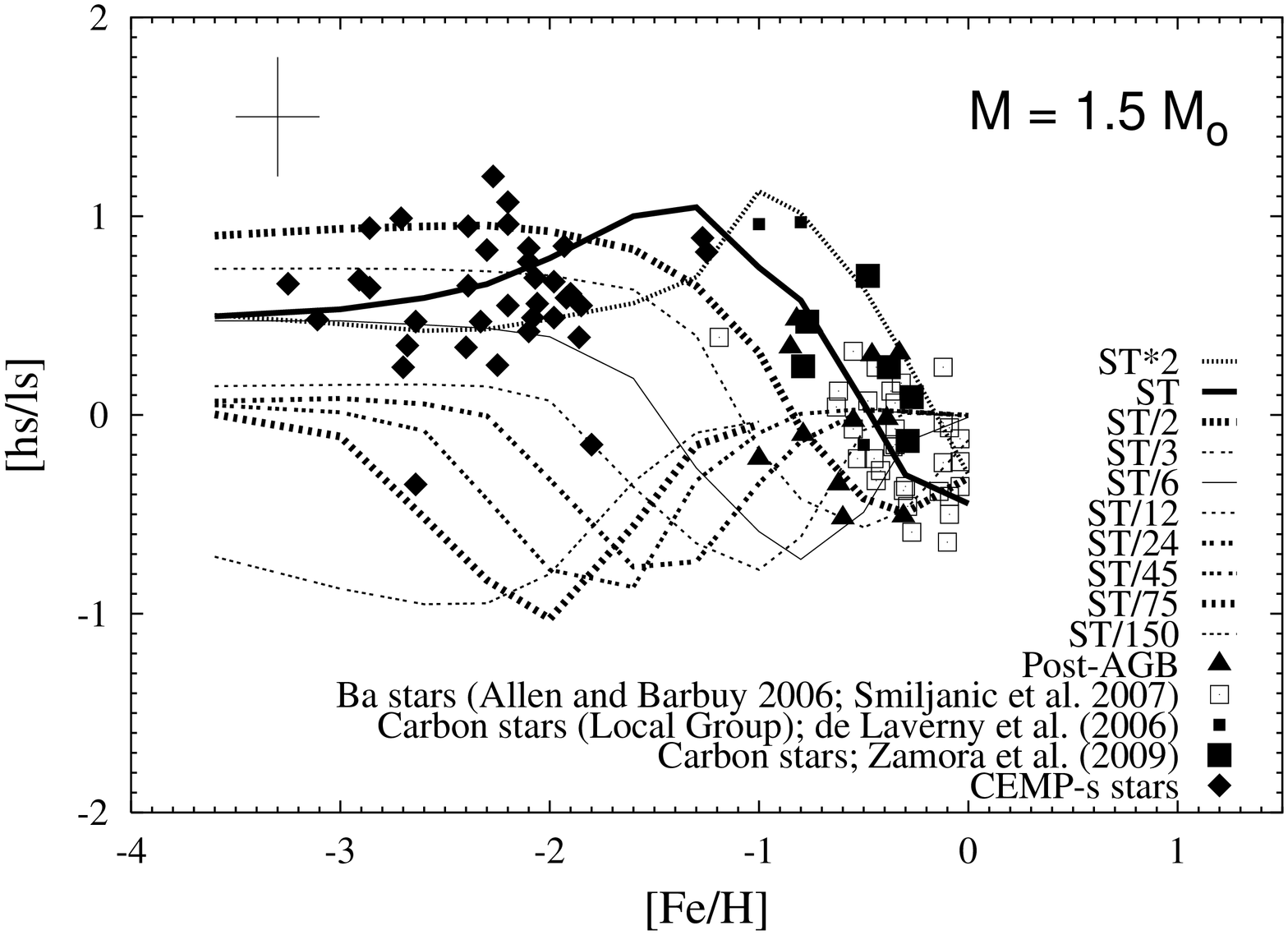}}
\resizebox{\hsize}{!}{\includegraphics[angle=-90]{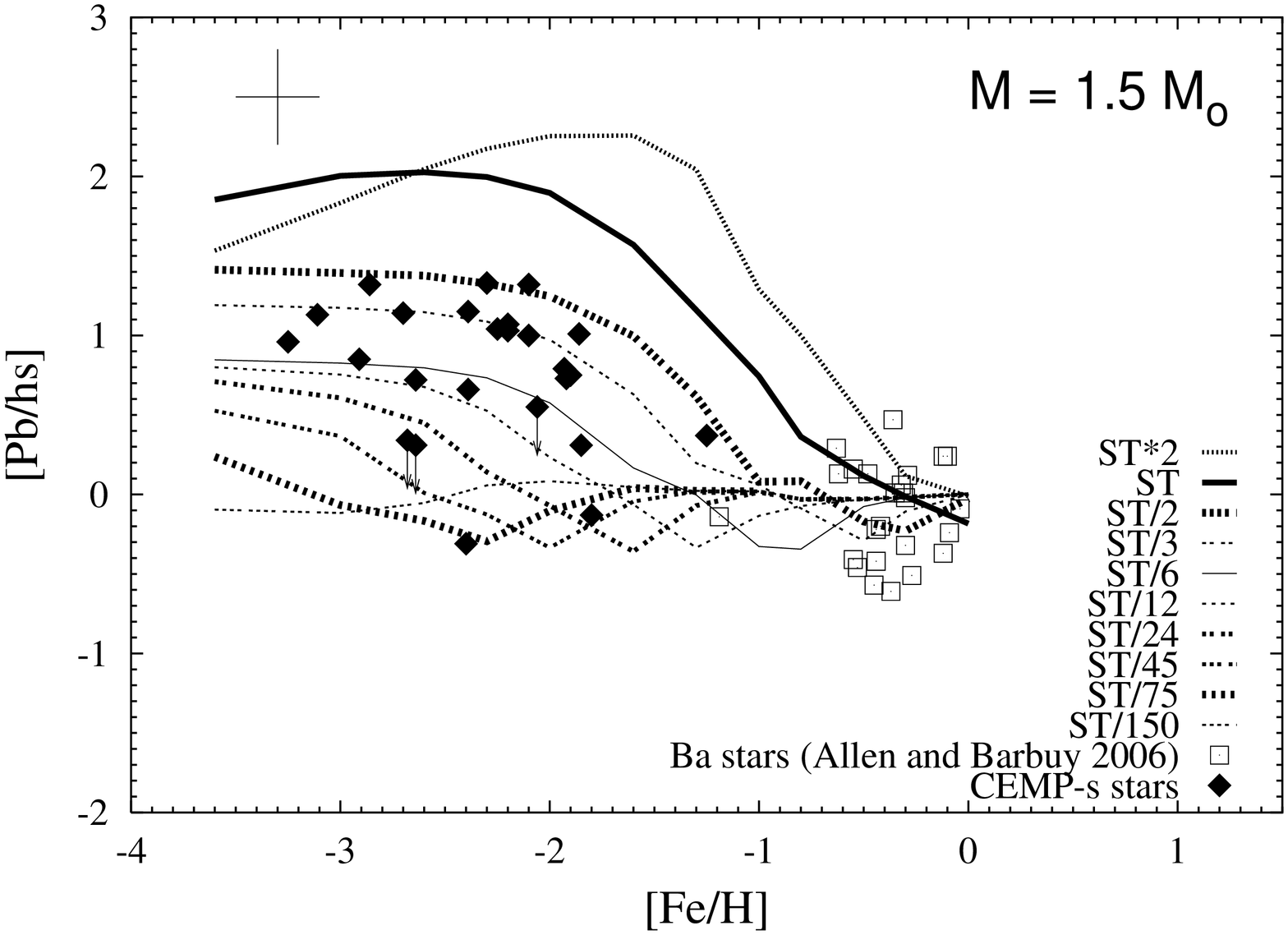}}
\caption{\footnotesize
Top panel. The first s-process indicator [hs/ls] for different
$^{13}$C-pocket  efficiencies. Comparison between theoretical models and
spectroscopic observations.
References are: Galactic Post-AGB (filled triangles; 
\citealt{reddy02}, \citealt{vanw00},  
\citealt{reyniers04,reyniers07}); barium dwarfs and giants  (open squares; 
\citealt{allen06,smil07}); metal-poor C(N) stars in the Local Group (small filled squares; 
\citealt{delaverny06};  Galactic C(N) stars (filled squares;
\citealt{zamora09}).
For CEMP-s stars (filled diamonds) references are \citet{aoki02}, 
\citet{aoki07}, 
\citet{barklem05}, 
\citet{cohen06}, 
\citet{goswami06},  
\citet{masseron06},        
\citet{pereira09},       
\citet{PS01},  
\citet{roederer08},        
\citet{thompson08},        
\citet{vaneck03}. 
}
\label{sindexes}
\end{figure}

Proton captures are now included in the post-process code to improve the
prediction
of Fluorine. Protons in the pocket and in the TP derive from $^{14}$N(n,
p)$^{14}$C.
Reaction rates of isotopes 
involving charged particles reactions involved in the various channels to
$^{19}$F are taken from the NACRE compilation \citep{angulo99}.
Different channels to $^{19}$F are: 	
\begin{itemize}
	\item $^{18}$O(n, $\gamma$)$^{19}$O($\beta$$^-$$\nu$)$^{19}$F 
	\item $^{15}$N($\alpha$, $\gamma$)$^{19}$F, where $^{15}$N derives 
         from $^{14}$N(n, $\gamma$)$^{15}$N and 
         $^{14}$C(n, $\gamma$)$^{15}$C($\beta$$^-$$\nu$)$^{15}$N
        \item $^{18}$O(p, $\gamma$)$^{19}$F
 	\item $^{18}$O(p, $\alpha$)$^{15}$N($\alpha$, $\gamma$)$^{19}$F
 	\item $^{13}$C in the ashes of the H-burning shell:
	$^{13}$C($\alpha$, n)$^{16}$O, $^{14}$N(n, p)$^{14}$C and
       $^{18}$O(p, $\alpha$)$^{15}$N in the early development of the TP,
        followed by $^{15}$N($\alpha$, $\gamma$)$^{19}$F   
\end{itemize}
For a recent discussion of the Fluorine production in AGB stars we refer
to \citet{lugaro04,lugaro08,abia09,cristallo09}.
Improvement of fluorine prediction, accounting for updated experimental
measurements of reaction rates, will be discussed elsewhere. Table~\ref{Ftest}
show the results of various tests to individuate the relative contribution
to $^{19}$F by  different nuclear
channels for an  AGB model of 2 \Msun,
[Fe/H] = $-$2.3 and case ST/6. Actually, $^{19}$F  production  does not
depend much on the $^{13}$C-pocket efficiency.

\begin{table*}
\caption{Tests for the origin of Fluorine.}
\label{Ftest}
\begin{center}
\begin{tabular}{lc}
\hline 
				Cases		 & {[F/Fe]} \\[0.5ex]
\hline
No channels                                     &   	0.00 \\
Only  $^{18}$O(n, $\gamma$)$^{19}$O($\beta$$^-$$\nu$)$^{19}$F                      &    1.18 \\
Only  $^{18}$O(p, $\gamma$)$^{19}$F                              &    3.47 \\
Only  $^{18}$O(p, $\alpha$)$^{15}$N                              &    3.50 \\
Only  $^{18}$O(p, $\alpha$)$^{15}$N and No $^{13}$C primary in the TP &    2.16 \\
All channels                              		& 3.80\\ [1.0ex]
\hline
\end{tabular}
\end{center}
\end{table*}

\begin{table*}
\caption{Fluorine and s-process productions for an AGB model of 2 \Msun;  [Fe/H] = --2.3; case ST/6.}
\label{pF}
\begin{center}
\begin{tabular}{lcccc}
\hline 
Cases      &    {[F/Fe]} &      {[Y/Fe]} & {[La/Fe]} & {[Pb/Fe]} \\ [0.5ex] 
\hline
No protons              &       1.65&    2.02  &   2.86  &    3.82 \\ 
With protons    & 3.54&         1.98    & 2.85  & 3.82  \\
With protons + primary $^{13}$C in the TP  &    3.80    &       2.01&
2.86&   3.81\\ [1.0ex]
\hline
\end{tabular}
\end{center}
\end{table*}

\begin{table*}
\caption{AGB model of 2 \Msun;  [Fe/H] = --2.3; s-process indexes for
different $^{13}$C-pocket efficiencies.}
\label{datam2}
\begin{center}
\begin{tabular}{lccccc}
\hline 
 Cases &		 {[ls/fe]}  &	{[hs/fe]}& 	{[Pb/Fe]} &	{[hs/ls]} &	{[Pb/hs]}\\
\hline
ST $\times$ 2     &  	1.68 &   		2.13  &    	4.21 &    		0.45    &		2.08 \\
ST       &  	1.36 &   		1.84  &    	4.12 &    		0.48    &		2.28 \\
ST/3     &  	1.55 &  		2.50  &    	4.00 &    		0.95    &		1.50 \\ 
ST/6     & 	 2.09  &  		2.82  &    	3.81 &    		0.73   & 		0.99 \\
ST/12    &	 2.41  &  		2.85  &    	3.42 &    		0.44   & 		0.57 \\
ST/24    & 	 2.43  &  		2.58  &    	2.65 &    		0.15   & 		0.07 \\
ST/45     &	2.20   & 		1.83    &  	1.72   & 		-0.37   	&	-0.11\\
ST/75     & 	1.74&   		1.07  &    	0.62 &   		-0.67   &		-0.45\\
ST/150    &	0.97  &  		0.14    &  	0.13   &		-0.83   &		-0.01\\ [1.0ex]
\hline
\end{tabular}
\end{center}
\end{table*}

\begin{figure}[t!]
\resizebox{\hsize}{!}{\includegraphics[angle=-90]{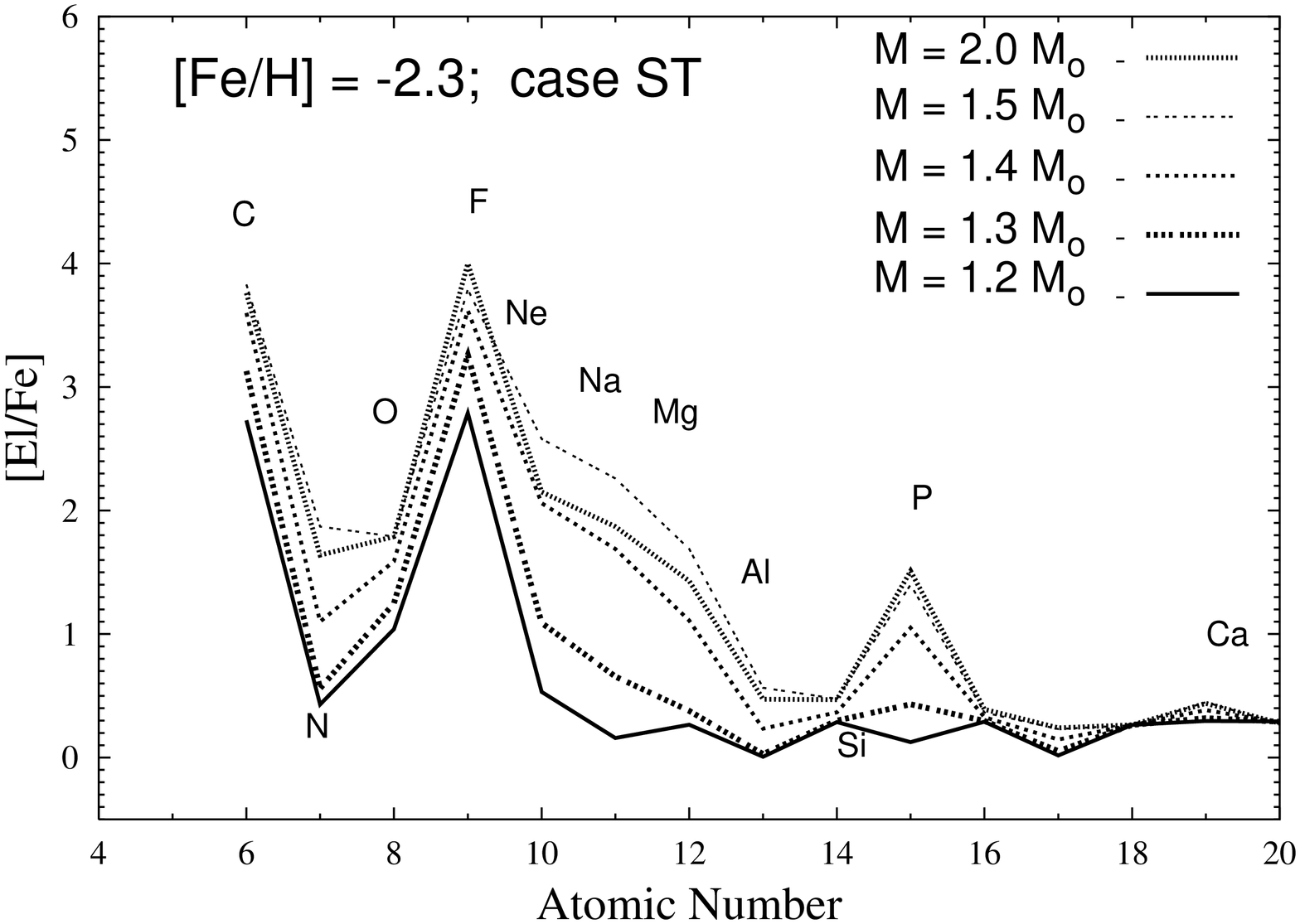}}
\resizebox{\hsize}{!}{\includegraphics[angle=-90]{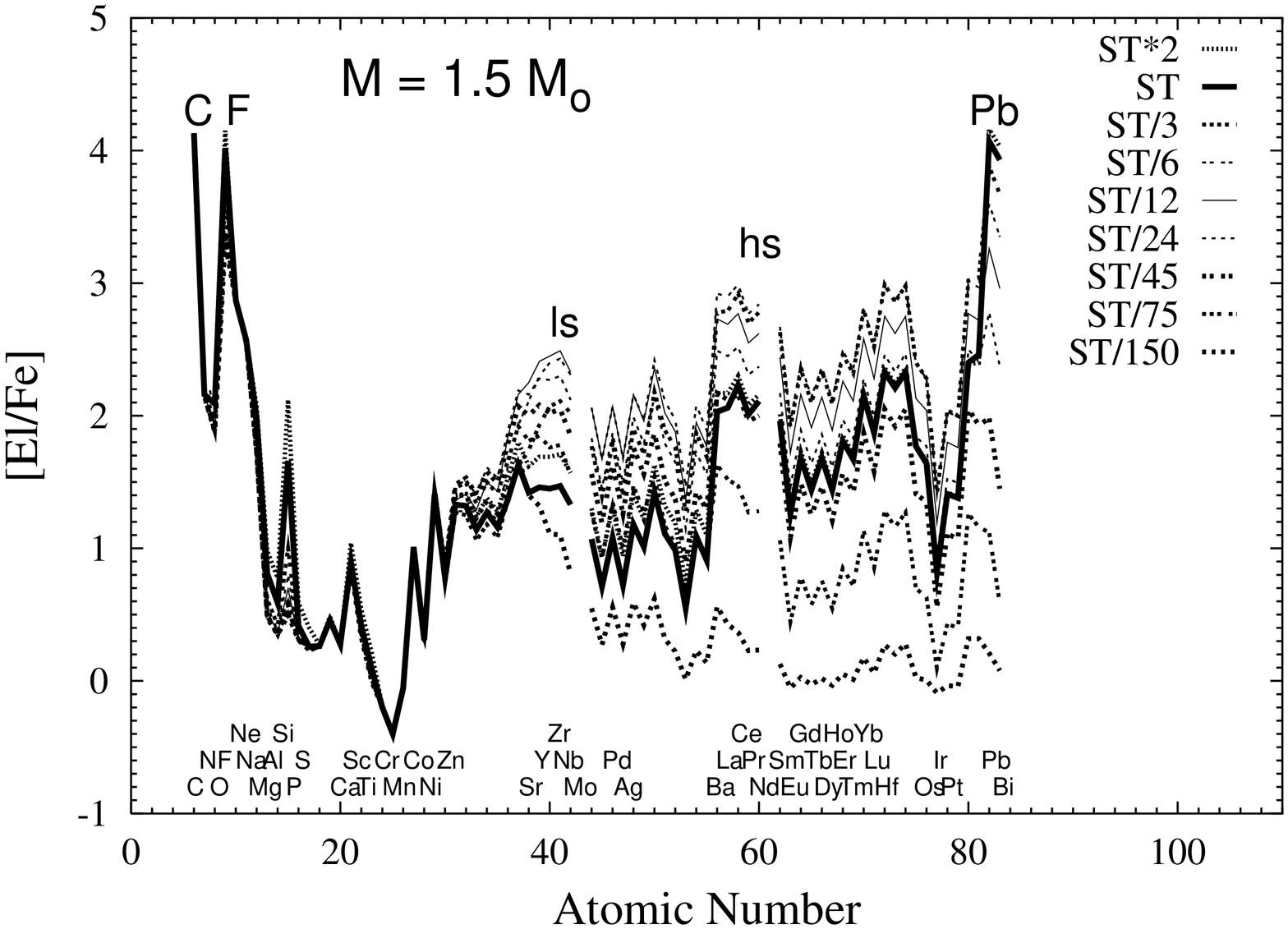}}
\caption{\footnotesize
(Top panel). Primary light elements expectation at [Fe/H] =
$-$2.6, case ST and different AGB initial masses. (Bottom panel). [El/Fe]
expectations as a function of the atomic number for AGB models of initial mass $M$ = 1.5 \Msun, metallicity
[Fe/H] = $-$2.6, and different $^{13}$C-pocket efficiencies.
}
\label{F}
\end{figure}

\subsection{Fluorine in CEMP-s stars}

Schuler et al. (2007), using the Phoenix spectrometer and Gemini-South
telescope discovered a CEMP-s star, HE 1305+0132, with [Fe/H] = $-$2.5
$\pm$0.5 and very enhanced Fluorine: [F/Fe] = 2.9. Subsequently, 
Schuler et al. (2008) with a higher resolution spectrum taken at the 
9.2 mt HET Telescope of the MacDonald
Observatory, better estimated [Fe/H] = $-$1.9, and derived [C/Fe] = 1.7, [N/Fe] = 1.5, 
[Ba/Fe] = 0.9. The star is a giant of \teff\ =  4462 $\pm$ 100 K; 
log $g$ = 0.80 $\pm$ 0.30. Because of  
the First Dredge Up (FDU), the C-rich and s-process-rich material transferred by 
the winds of the primary AGB was diluted by 1 dex or more over the convective envelope
of the observed star.
Our typical AGB predictions would fit these data once 1 dex dilution is
applied, with a further prediction of [Pb/Fe] $\sim$ 2.5. However,
a higher spectroscopic resolution is necessary for a definite confirmation
and for detection of other heavy neutron-capture elements besides Ba.

\begin{figure}[t!]
\resizebox{\hsize}{!}{\includegraphics[angle=-90]{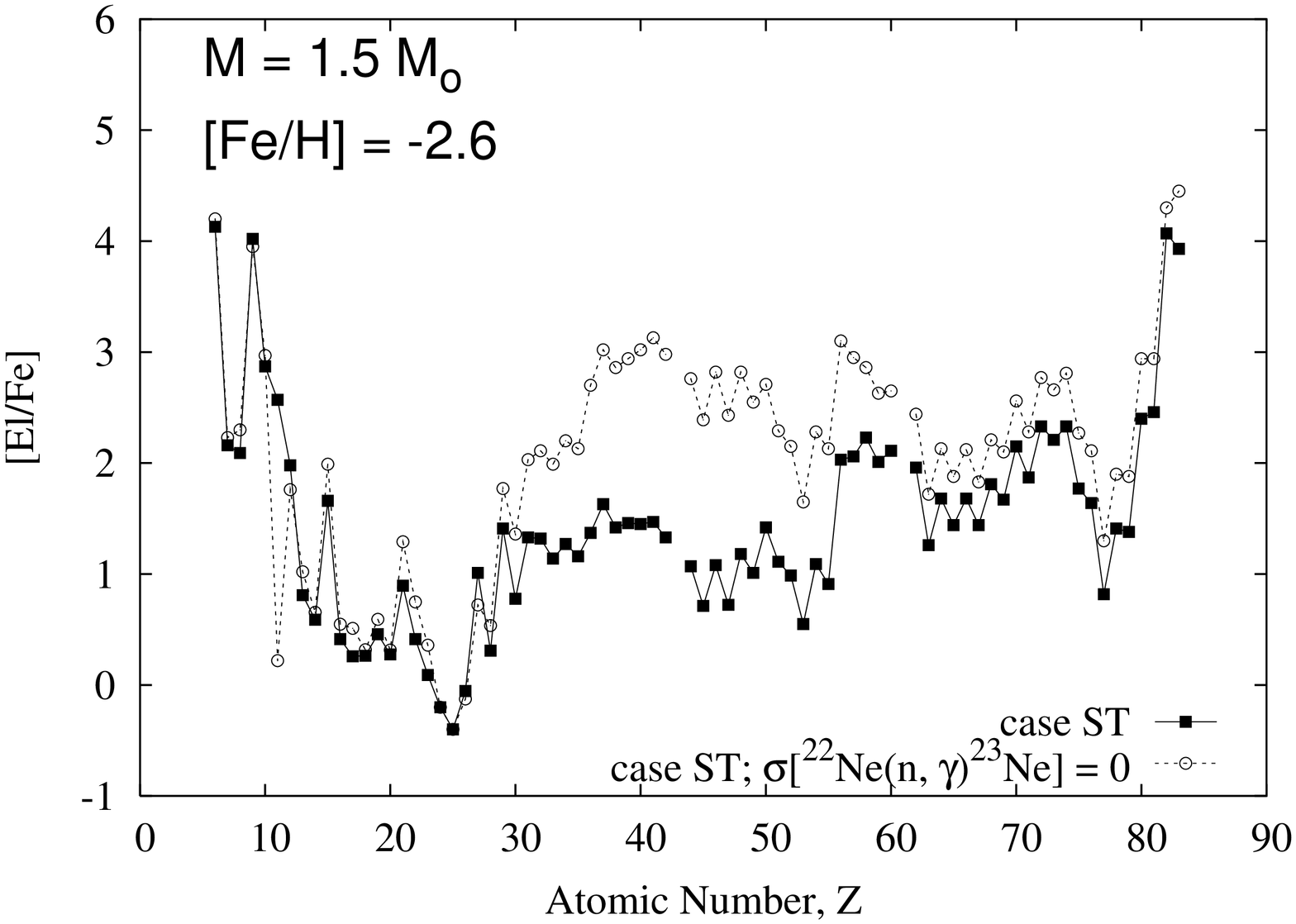}}
\resizebox{\hsize}{!}{\includegraphics[angle=-90]{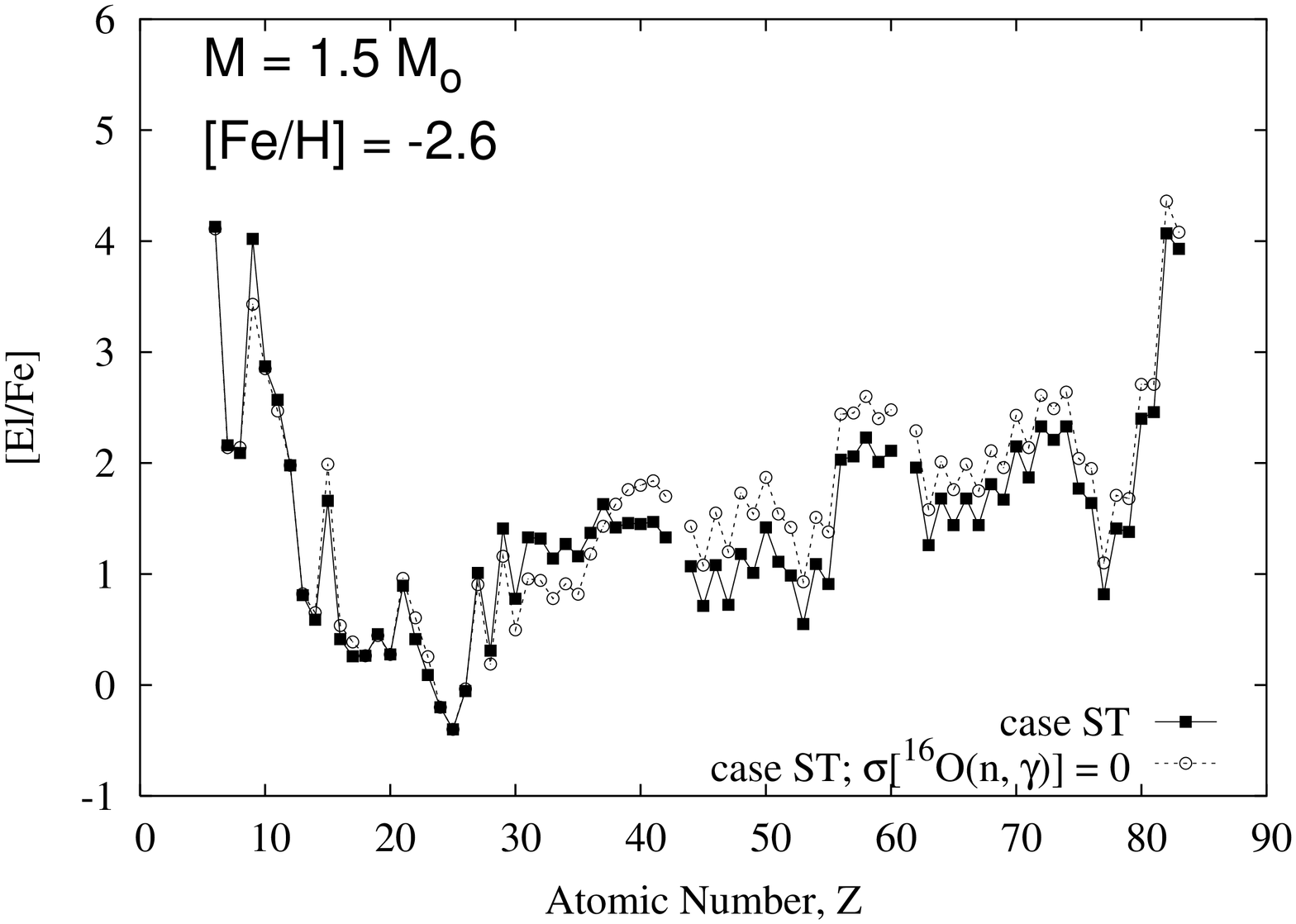}}
%\resizebox{\hsize}{!}{\includegraphics[angle=-90]{XTW_AAbab9ltHTp1p5m1p5z5m5_nosigc12nglt_n5_bw.ps}}
\caption{\footnotesize
The neutron poisoning effect of primary $^{22}$Ne (top panel) and of primary
$^{16}$O (bottom panel) for an AGB model of
[Fe/H] = $-$2.6, initial mass $M$ = 1.5 \Msunb and case ST.  
}
\label{figpoisons}
\end{figure}

\section{Primary light elements production.}
Together with Carbon and Fluorine, a number of light elements are
synthesised in a primary way in low mass AGB stars of Halo metallicity. This includes Ne
(in the form of $^{22}$Ne) and its neutron capture progenies
$^{23}$Na. Besides, a very large overabundance of the $^{25}$Mg
and $^{26}$Mg isotopes are obtained, by partial depletion of $^{22}$Ne  by $\alpha$ captures
in the convective TP. Some primary $^{16}$O is  also produced by $\alpha$
captures on $^{12}$C during the TP.
Figure ~\ref{F} (top panel) shows the envelope AGB predictions at the last
TDU episode  of light elements for AGB models of metallicity [Fe/H] = $-$2.3, 
case ST and initial masses 2.0, 1.5. 1.4, 1.3 and 1.2 \Msun. The general increase of the
production factors with the increase of the initial mass is a direct
consequence of the increase of the cumulative He-instershell mass mixed with
the envelope, which grows with the number of TDU episodes.
The primary contribution to Nitrogen in the envelope was made
in the inactivated H-shell that is also mixed with the envelope at
each TDU episode.  
When comparing the C/N ratio observed in CEMP-s stars, one should consider
that an important fraction of $^{12}$C is possibly converted to $^{14}$N by
extra-mixing episodes in the observed star. The effect on the s-process by
neutron capture of primary $^{16}$O and $^{22}$Ne is illustrated in
Figure~\ref{figpoisons}. 

\section{Conclusions} 
For Halo metallicities, the $^{16}$O bug we had in the $^{13}$C-pocket,
which consisted in discarding neutron capture on the $^{16}$O isotopes released by 
$^{13}$C($\alpha$, n)$^{16}$O, affects 
[ls/Fe], [hs/Fe], [Pb/Fe] in an important way for the highest 
choices of the $^{13}$C-pocket efficiency (cases ST $\times$ 2 and ST).
Instead, the s-process indexes [hs/ls], [Pb/hs] are only marginally 
affected.
For Galactic disc AGBs, the bug effect is negligible for all
$^{13}$C-pocket choices. 
Introduction of proton captures in the code does not affect the 
s-process abundances. However, it boosts  the prediction of 
Fluorine in very metal-poor s-enhanced stars. A primary production of F, comparable with the C
enhancement, is predicted in AGB stars of low metallicity.
The only spectroscopic observation of Fluorine in a CEMP-s star,
is the giant HE 1305+0132 by Schuler et al. (2007, 2008), and is quite in agreement with our F prediction. 
Higher resolution spectra are  however necessary for a definite
confirmation.

\begin{acknowledgements}
We are indebted to Franz K{\"a}ppeler for his continuous 
support in implementing the experimental neutron capture  measurements.
R.G. deeply acknowledges for local support the Centre for Stellar and Planetary
Astrophysics, School of Mathematical Sciences of the Monash University
(Victoria, Australia) and the Organisers of the Workshop for a marvellous
permanence at the University of Canterbury, Christchurch (New Zealand). 
\end{acknowledgements}

\bibliographystyle{aa}

\end{document}